# A CT Image Classification Network Framework for Lung Tumors Based on Pre-trained MobileNetV2 Model and Transfer learning, And Its Application and Market Analysis in the Medical field

Ziyang Gao[1,3], Yong Tian[1,4], Shih-Chi Lin[2,5], Junghua Lin[3,6]

[1]School of Engineering,China University of Petroleum-Beijing at Karamay Campus,Xinjiang,China
[2]Epocrates, Athenahealth Inc. , Austin, TX, USA
[3]Suffolk University, Boston, USA

[3]2022015929@st.cupk.edu.cn
[4]2022015945@st.cupk.edu.cn
[5]oslo.lin@alumni.usc.edu
[6]rjl00774@su.suffolk.edu

**Abstract.** In the medical field, accurate diagnosis of lung cancer is crucial for treatment. Traditional manual analysis methods have significant limitations in terms of accuracy and efficiency. To address this issue, this paper proposes a deep learning network framework based on the pre-trained MobileNetV2 model, initialized with weights from the ImageNet-1K dataset (version 2). The last layer of the model (the fully connected layer) is replaced with a new fully connected layer, and a softmax activation function is added to efficiently classify three types of lung cancer CT scan images. Experimental results show that the model achieves an accuracy of 99.6% on the test set, with significant improvements in feature extraction compared to traditional models.With the rapid development of artificial intelligence technologies, deep learning applications in medical image processing are bringing revolutionary changes to the healthcare industry. AI-based lung cancer detection systems can significantly improve diagnostic efficiency, reduce the workload of doctors, and occupy an important position in the global healthcare market. The potential of AI to improve diagnostic accuracy, reduce medical costs, and promote precision medicine will have a profound impact on the future development of the healthcare industry.

**Keywords:** Image Classification, Transfer Learning, MobileNetV2, Commercial Application of Medical Products

## 1. Introduction

Lung tumors, particularly lung cancer, represent a significant global health challenge, as they are among the leading causes of cancer-related deaths. Early diagnosis is crucial for improving patient survival rates. With rapid advancements in medical imaging technology, CT scans have become a mainstream method for diagnosing lung tumors, playing a vital role in distinguishing tumor types and monitoring the spread of lesions. However, due to the complexity of lung structures and the diversity of tumor types, traditional manual analysis methods face limitations in both efficiency and accuracy. Thus, the application of computer science and deep learning algorithms to automatically classify CT scan images has become a popular research direction in medical image analysis [1].

The development of machine learning has made computer-aided diagnosis and intelligent healthcare a reality. Convolutional Neural Networks (CNNs) have gained significant attention in deep learning due to their excellent performance in image recognition and classification tasks. CNNs can automatically learn data features, mapping raw input images to final classifications. MobileNetV2, a newer version of the MobileNet architecture, is particularly well-suited for resource-constrained environments such as mobile and embedded systems. By employing depthwise separable convolutions and the inverted residual structure, MobileNetV2 significantly improves computational efficiency without sacrificing accuracy, making it an ideal candidate for medical image classification.

Medical images are inherently complex due to factors such as lesion overlap, background noise, and low resolution. The classification results directly impact disease diagnosis and treatment, requiring high precision and accuracy. However, challenges such as small standardized image datasets, overfitting, and low generalization ability hinder the performance of current deep learning models [2]. Therefore, responding to the uneven distribution of global healthcare resources and addressing the need for a shared medical future, it is crucial to propose a customized medical image classification network tailored for specific organs, such as lung tumors.The medical AI market, particularly in image analysis for lung disease detection, is rapidly growing. AI's application in medical image classification, especially for lung tumors, holds immense potential in improving diagnostic efficiency, reducing human error, and lowering healthcare costs. AI technologies are expected to transform healthcare, offering significant commercial opportunities while advancing global health outcomes.

To address these limitations, this paper presents a transfer learning network based on the pre-trained MobileNetV2 model for lung tumor CT image classification. Our contributions focus on three key aspects: First, we utilize the pre-trained MobileNetV2 model and fine-tune it by replacing the fully connected layer and adding a softmax activation function to adapt it for lung CT image classification. Second, we design a complete data preprocessing pipeline, including image conversion and normalization, to improve the model's generalization ability. Finally, we conduct a detailed performance evaluation, assessing not only accuracy but also precision, recall, and F1 score to comprehensively evaluate the model's effectiveness.

## 2. Related Work

In recent years, significant progress has been made in the field of classification or segmentation of medical images. Through various machine learning algorithms represented by deep learning, researchers have significantly improved the accuracy of classifying or segmenting various medical images. These advances not only improve the accuracy of diagnosis and the efficiency of medical resource utilization, but also make the classification and segmentation process more intelligent and automated. In addition, the CNN-ResNet-Transfer Learning method has been widely used in remote sensing analysis, autonomous driving, cyberspace security and other fields. The following is an overview of the development of this field:

Alex Noel Joseph Raj [3] put forward a kind of based on MKFCM clustering technology and orthogonal learning particle swarm optimization (OLPSO) and hybrid classifier (a combination of the SVM and ANN) method of medical image segmentation and classification from CT scans of the lung in the image segmentation and classification of tuberculosis. In-depth analysis of the experiment reveals that the proposed method can effectively identify and classify tuberculosis from CT scan images. The sensitivity was 89.87%, the specificity was 82.88%, the positive likelihood ratio was 5.24, the negative likelihood ratio was 0.122, the prevalence rate was 7.69%, and the false positive rate was significantly reduced.

The traditional machine learning method with its solid theoretical basis and validity through the test of time, has been widely used in many areas, however, with the rapid development of deep learning, researchers began to tend to use more depth convolution neural network for classification and segmentation of the image.

M. a. h. Akhand, Shuvendu Roy, Nazmul Siddique and others [4] learn by migration (TL) and pipelining fine-tuning strategy, puts forward A convolutional neural network (DCNN) based on the depth of facial emotion recognition (FER) method. The method is tested on KDEF and JAFFE databases using eight pre-trained DCNN models. Namely VGG-16, VGG-19, ResNet-18, ResNet-34, ResNet-50, ResNet-152, Inception-v3, and DenseNet-161. DenseNet-161 is used to perform 10-fold cross validation on the data set, and the prediction accuracy reaches 96.51% and 99.52% respectively. Experiments show that the proposed method has significant advantages over the existing facial emotion recognition methods.

Akhilesh Kumar Gangwar and Vadlamani Ravi [5] built a Inception - ResNet - v2 model based on preliminary training of migration study hybrid model, used to detect diabetic retinopathy. They tested and evaluated the performance of the proposed model on Messidor-1 and APTOS datasets, with accuracies of 72.33% and 82.18%, respectively. Compared with GoogleNet, the accuracy is improved by 6.3%. Experimental data reveal that the proposed model outperforms the published conventional models.

Jiaqi Hu, Yanqiu Zou, and Biao Sun et al. [6] used the method of convolutional neural network transfer learning to construct three one-dimensional convolutional neural network models, namely CNN-1D, Resnet-1D, and Inception-1D, and applied them to Raman spectroscopy analysis related to pesticide detection. The comparative evaluation of the three models with or without transfer learning is carried out to determine the accuracy of the method: the accuracy of the three models reaches 98.38%-99.88%, and the accuracy of spectral classification is significantly improved, which is 6%, 2% and 3% respectively. The experimental results show that the feature extraction ability and generalization ability of Raman spectroscopy model can be improved by transfer learning, so that the application range of Raman spectroscopy is expanded and the computational cost of training is reduced.

**3. Method**

MobileNetV2 [7] addresses the trade-off between computational efficiency and model complexity by introducing depthwise separable convolutions and inverted residual blocks. This design reduces computational load while maintaining model expressiveness. Compared to traditional deep networks like VGG, MobileNetV2 is more lightweight and efficient, making it ideal for resource-constrained environments. Its adaptability through transfer learning allows for fine-tuning on medical image tasks, such as lung tumor classification, enhancing diagnostic accuracy. MobileNetV2's efficiency and performance make it a powerful tool in medical image analysis, driving advancements in AI-assisted healthcare while reducing computational costs [8].

*3.1 Diagnosis model for pulmonary lesions*

In this paper, a novel deep learning network framework based on MobileNetV2 and transfer learning is proposed to adapt to a specific pneumonia CT image classification task by fine-tuning.

**Table1 .** Diagnosis Model Architecture

| Layer |
| --- |
| MobileNetV2 |
| GlobalAveragePooling2D |
| Dense |
| Dropout |
| Dense (softmax) |

*3.2 Model training process*

In this paper, we use a pre-trained MobileNetV2 model with weights trained on the ImageNet-1K dataset (version 2). MobileNetV2, a lightweight network designed for mobile devices, introduces inverted residual blocks and depthwise separable convolutions to reduce computational cost while maintaining high efficiency. We replace the final fully connected layer with a new layer tailored to the specific output size, and add a softmax activation function to provide a probability distribution for predictions. We configure the model by selecting the device (GPU or CPU) using the .to(device) method, define the cross-entropy loss function to measure prediction errors, and implement the StepLR learning rate scheduler to adjust the learning rate during training. To update the weights of the model, we create a Stochastic Gradient Descent (SGD) optimizer, which is calculated as follows:

$$\theta_{t+1} = \theta_t - \alpha \nabla f_i(\theta_t)$$

Where: $\theta$ denotes the model parameters; $t$ denotes the current number of iterations; $\alpha$ represents the learning rate and controls the step size; $f_i$ represents the loss function for the i-data point or batch

The model undergoes 10 epochs of training with a cyclic approach. During each epoch, we accumulate loss, accuracy, and sample counts for each batch, using backpropagation to update model parameters. For the validation set, loss and accuracy are calculated without gradients, and validation accuracy serves as the model evaluation criterion. The best model is saved using deepcopy(), and the learning rate scheduler is adjusted to enhance convergence. Table 2 shows the model's performance across 10 epochs. As iterations increase, the accuracy typically improves, and the loss function decreases. After 10 epochs, Train Loss and Validation Loss reach approximately 0.1649 and 0.2096, respectively, with Train Accuracy and Validation Accuracy both achieving 1.

**Table2.** The accuracy and loss function of the model on the training set and validation set within 10 epochs

| Epoch | Train Loss | Train Accuracy | Validation Loss | Validation Accuracy |
|---|---|---|---|---|
| 1 | 0.2211 | 0.9402 | 0.2836 | 0.8939 |
| 2 | 0.1958 | 0.9482 | 0.2593 | 0.9242 |
| 3 | 0.1912 | 0.9562 | 0.2436 | 0.9242 |
| …… | …… | …… | …… | …… |
| 8 | 0.1666 | 0.9721 | 0.2218 | 0.9242 |
| 9 | 0.1920 | 0.9323 | 0.2189 | 0.9242 |
| 10 | 0.1649 | 0.9522 | 0.2096 | 0.9394 |

The Accuracy and loss functions of the model on the training set and validation set are shown as follows. The Loss log and Accuracy log curves of the training set and validation set fit well in general in figure 1.

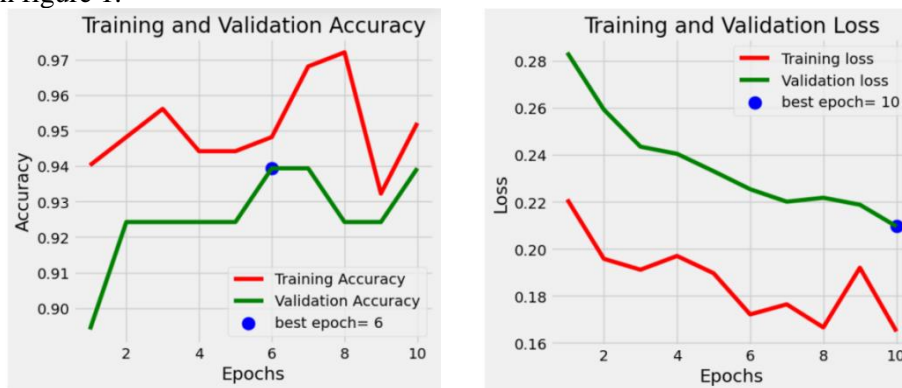

**Figure 1.** The accuracy and loss function of the model on the training set and validation set (After fine-tuning)

## 4. Results

*4.1. Datasets*
The dataset is divided into two main directories: train and test, with three subdirectories for each category: Normal, Pneumonia, and Lung Tumor. These subdirectories store CT scans for training, validation, and testing. LabelEncoder is used to integer-code the text labels, and the test set is randomly shuffled. The training set contains 80% of the original data, while the validation set holds 20%. A class called Pipeline is defined to efficiently load and preprocess the data. The transforms module is used for image preprocessing to fit the pre-trained MobileNetV2 model, and two data loaders are created for training and validation.

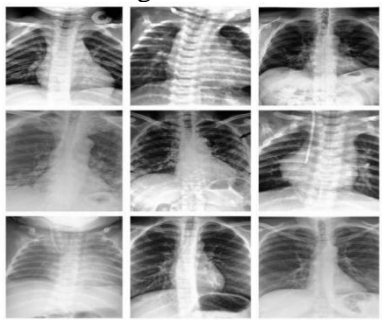

**Figure 2.** Part of the pneumonia CT scan images in the dataset

*4.2. Classification Results and Analysis of the Model*

The function predict is defined to reshape the image into a batch size of 1 tensor for predicting a single image in the test set. Make a prediction using the best model and return the predicted class index and probability. We iterate over the test set and collect the true labels, predicted results, and predicted probabilities into the corresponding lists for evaluating the performance of the model.

To measure the classification performance of the model, the scikit-learn library is invoked to generate a classification report that provides classification metrics, such as Precision, Recall, F1-score, and so on, whose mathematical expressions are as follows:

$$\text{Precision} = \frac{TP}{TP+FP} \qquad (1)$$

$$\text{Recall} = \frac{TP}{TP+FN} \qquad (2)$$

$$\text{F1-score} = \frac{2 \times \text{Precision} \times \text{Recall}}{\text{Precision} + \text{Recall}} \qquad (3)$$

It is worth noting that this model overcomes the shortcomings of general models that cannot have high precision and high recall at the same time, and improves both of them to more than 0.952 at the same time, indicating that this model can balance the prediction of positive and negative categories, has good robustness and discrimination ability, and performs well in feature engineering and sample balancing. It can maintain high performance in different situations in Table 3.

**Table 3.** A classification report that evaluates the classification performance of the model

|  | precision | recall | f1-score | support |
|---|---|---|---|---|
| 0 | 1.00 | 1.00 | 1.00 |  |
| 1 | 0.94 | 0.97 | 0.95 |  |
| 2 | 0.96 | 0.96 | 0.96 |  |
|  |  |  |  |  |
| accuracy | 0.952 | 0.975 | 0.963 | 251 |

**5. Conclusion**

In this paper, a deep convolutional neural network based on MobileNetV2, utilizing transfer learning, is proposed for classifying three types of lung CT scans: Normal, Pneumonia, and Lung Tumor. The original MobileNetV2 fully connected layer is replaced with a new one, followed by adding a softmax activation function and custom classification head. After 10 iterations, the training set loss reaches 0.1649, and the validation loss reaches 0.2096. The final model achieves 95.2% accuracy on the test set, demonstrating outstanding classification performance.

The results show that the proposed network model effectively classifies the three types of lung CT images. While potential issues such as limited defense against adversarial attacks and accuracy fluctuations may still exist, this model provides valuable insights for a customized medical image classification network used for disease diagnosis and monitoring. This model not only improves the accuracy of lung disease diagnosis but also supports early screening of lung tumors and other diseases, contributing to the development of intelligent healthcare.

From a business and market analysis perspective, the continuous development of medical image automation, particularly in the context of deep learning, presents a vast potential for growth. The application of transfer learning using pre-trained models can significantly reduce the costs associated with data labeling and model training, improving diagnostic efficiency in hospitals, clinics, and health management centers. This will help reduce the risks and errors of manual diagnosis and push forward the trend of automation in the healthcare industry.

Future work could focus on integrating this model with other neural networks, such as Inception, DenseNet, and Xception, to further enhance the accuracy of pneumonia CT image classification. Additionally, the model could be extended to classify other diseases such as brain and gastric conditions. Expanding datasets for model training will also help fine-tune the model and reduce overfitting, thus fostering broader clinical applications.